\documentclass{article}
\usepackage{amsmath}
\usepackage{graphicx}
\def\p{\partial}

\def\g{\gamma}

\def\de{\delta}
\def\D{\Delta}
\def\De{\Delta}
\def\ov{\overline}
\def\ld{\lambda}
\def\Ld{\Lambda}

\def\om{\omega}
\def\Om{\Omega}

\def\a{\alpha}

\def\pdellx'{\frac{\partial}{\partial x'}}
\def\pdellw'{\frac{\partial}{\partial w'}}
\newcommand{\be}{\begin{equation}}
\newcommand{\ee}{\end{equation}}
\def\bed{\begin{displaymath}}
\def\eed{\end{displaymath}}
\def\bea{\begin{eqnarray}}
\def\eea{\end{eqncrray}}
\def\[{$$}
\def\]{$$}
\begin{document}
\title{A Confining Model for Charmonium and\\ New Gauge Invariant Field Equations}
\author{Jong-Ping Hsu\footnote{jhsu@umassd.edu}\\
Department of Physics, \\
University of Massachusetts Dartmouth, \\
North
Dartmouth, MA 02747-2300, USA}

\date{\today}
\maketitle
{\small  We discuss a confining model for charmonium in which the attractive force are derived from a new type of gauge field equation with a generalized $SU_3$ gauge symmetry.  The new gauge transformations involve non-integrable phase factors with  vector gauge functions $\om^a_{\mu}(x)$.  These transformations reduce to the usual $SU_3$ gauge transformations in the special case $\om^a_\mu(x) = \p_\mu \xi^a(x)$.  Such a generalized  gauge symmetry leads to the fourth-order equations for new gauge fields and to the linear confining potentials.  The fourth-order field equation implies that the corresponding massless gauge boson has non-definite energy.  However, the new gauge boson is permanently confined in a quark system by the linear potential.   We use the empirical potentials of the Cornell group for charmonium to obtain the coupling strength $f^2/(4\pi) \approx 0.19$ for the strong interaction.  Such a confining model of quark dynamics could be compatible with perturbation.  The model can be applied to other quark-antiquark  systems. }
\bigskip

\section{Introduction }

The excited states of charmonium provide information regarding forces and potentials between charmed quarks or c-quarks\cite{1}.  The forces between charmed quarks come from a linear potential  $\propto r$ and a Coulomb-type potential $\propto  r^{-1}$. It is interesting to note that the coupling strength associated with the Coulomb-type potential is not purely electromagnetic.  This suggests that the color $SU_3$ gauge fields, which generate a linear confinement potential for c-quarks, must also produce a Coulomb-type potential.  Thus, it is desirable from a physical viewpoint that one investigates a model of charmonium with new gauge fields, which have these properties explicitly.  

In this paper, we suggest that a new type of gauge invariant field equation involving fourth-order space-time derivatives could lead to these required properties.  We show that these potentials between a c-quarks and anti-c-quark  in a charmonium can be obtained by assuming a generalized gauge symmetry for the conservations of quark current.  The ideas of generalization are (i) to replace the usual (Lorentz) scalar gauge function $\xi^a(x)$ by a vector function $\om^a_{\mu}(x)$, and (ii) to replace the phase factor in usual gauge theory by a non-integrable phase factor, which contains the vector gauge function $\om^a_{\mu}(x)$\cite{2}.    In the special case when the vector gauge function $\om^a_{\mu}(x)$ can be expressed as the space-time derivative of an arbitrary scalar function $\xi^a(x)$, i.e. $\om^a_\mu(x) = \p_{\mu}\xi^a(x)$, then the non-integrable phase factor reduces to the usual phase factor.  In this way, the generalized $SU_3$ gauge transformations becomes the usual $SU_3$ gauge transformations.  In the literature, Yang stressed that electromagnetism is the gauge invariant manifestation of the non-integrable phase factor $exp(ie\oint  A_{\mu} dx^{\mu})$, which provides an intrinsic and complete description of electromagnetism\cite{3}.   We used it to explore forms of gauge fields, wrapping numbers and quantization conditions in gauge field theories\cite{4}.
  
 We argue that there is a physical justification for the fourth-order gauge field equation in particle physics based on the  principle of  generalized gauge symmetry and the experimental non-observability of a single quark.  The new gauge bosons associated with the fourth-order equation have non-definite energies, otherwise there is no essential difficulty, according to Pais and Uhlenbeck, and others\cite{5,6,7}.  We note that such a negative energy of the new gauge boson will not upset the stability of a quark system at quantum level and will not contradict experiment.  The reason is that the quarks satisfy the Dirac equations and, hence, a system of quarks has the ground state.  Furthermore, the gauge boson with negative energy is permanently confined in quark systems by a linear potential.  Thus, the new gauge boson will not lead to observable negative energies in the physical world.

\section{A Generalized  $U_{1}$ and $SU_3$ Gauge Symmetries}

For clarity and comparison, let us consider simultaneously the generalized gauge symmetries related to $U_1$ and  $SU_3$ groups.  Suppose gauge fields $B_\mu(x)$, and  fermion $\psi(x)$ are associated $U_1$, while gauge fields $H^a_\mu (x)$ and quark field $q(x)$ are associated with $SU_3$.   The new gauge transformations for $B_\mu(x)$ and $H^a_\mu(x)$ are assumed to involve  vector gauge functions $\Ld_\mu(x)$ and $\om^a_\mu(x)$,
\be
B'_\mu(x) = B_\mu(x) + \Ld_\mu (x),  \ \ \ \  for  \ \   U_1,
\ee
%%%%%%%%%15%%%1
 \be
H'_\mu(x) = H_\mu(x) + \om_\mu (x) - if\int^{x}dx'^\ld[ \om_{\ld}(x') , H_\mu(x)],     \ \ \  for   \ \  SU_3,
\ee
$$
   H_\ld(x)=H^a_{\ld}(x)L^a, \ \ \ \  \om_\ld(x)=\om^a_{\ld}(x)L^a,   \ \ \ \  [L^a , L^b]= if^{abc} L^c, \ \ \ \   L^a=\frac{\ld^a}{2},
$$
%%%%%21%%%2554%%%37%%%35%%%512%%%2
where $a, b, c=1,2,....8$ are  indices of the color $SU_3$ group, which has 8 generators $L^a$, $f^{abc}$ are structure constants, and $\ld^a$ are Gell-Mann matrices\cite{8}.  For fermions and quarks, the new gauge transformations are
\be
\psi'(x) =   \Om(x)\psi(x), \ \ \ \ \ \  \ov{\psi}'(x) = \ov{\psi}(x)  \Om(x)^{-1}, \ \ \ \ \ \  U_1,
\ee
%17%%%3
\be
q'(x) = \Om_{\om}(x) q(x), \ \ \ \ \   \ov{q'}(x) =\ov{q}(x)\Om_{\om}^{-1},    \ \ \ \ \ \ \ \ \ \ \   SU_3,
\ee
%%4
\be
   \Om(x)= exp\left(- ig_b \int^{x} \Ld_\ld(x') dx'^\ld \right),  
\ee
%18$$$4%%%3%6%%%%%%%5
\be
 \Om_{\om}(x)=exp\left[- i f \int^{x}  \om_{\ld}(x') dx'^\ld \right]  = \left[ 1 - i f \int^{x}  \om^a_\ld (x') dx'^\ld L^a \right] ,
\ee
%%7%%%6
where $\om^a_\ld (x)$  is an infinitesimal (Lorentz) vector gauge function.  
  The paths in (5) and (6)  could be arbitrary, as long as they end at the point $x\equiv x^\nu$.  
As usual, the  gauge covariant derivatives are defined as
\be
   \De_{b\mu}= \p_\mu + ig_b B_\mu(x), \ \ \ \ \ \ \    U_1,
\ee
%%%%%%%5%%25%%24%%%10%%%6%%%%10%%%7
\be
\De_\mu=\p_\mu + if H^a_\mu (x) L^a,  \ \ \ \ \ \   SU_3.
\ee
%%%%%11%%%%8
The $U_{1}$ and $SU_3$ gauge curvatures are defined as usual,
\be
[\De_{b\mu}, \D_{b\nu}]=ig_b B_{\mu\nu}, \ \ \ \ \ \ \ \    U_1,  
\ee
%%%%%6%%%26%%25%%%%11%%%7%%%12%%%9
\be
[\De_{\mu} , \D_\nu]=if H_{\mu\nu}^a L^a,  \ \ \ \ \ \ \ \ \ \   SU_3,
\ee
%%%%%%15%%%%13%%%%10
where
$$
B_{\mu\nu} = \p_\mu B_\nu - \p_\nu B_\mu,
$$
\be
H^a_{\mu\nu}(x) = \p_\mu H^a_\nu(x) - \p_\nu H^a_\mu(x) - f f^{abc} H^b_\mu(x)H^c_\nu(x),
\ee
%%%%%28%%532%%%%%59%%557%%%%%40%%%38%%%16%%%14%%%11
or
$$
H_{\mu\nu}(x) = \p_\mu H_\nu(x) - \p_\nu H_\mu(x) + i f [H_\mu(x), H_\nu(x)].
$$
We have the following new gauge transformations for $B_{\mu\nu}(x)$, $\p^\mu B_{\mu\nu}(x)$,  $H_{\mu\nu}(x) =H^a_{\mu\nu}(x)L^a $, and so on:
\be
B'_{\mu\nu}(x)= B_{\mu\nu}(x) +\p_\mu \Ld_\nu (x) - \p_\nu \Ld_\mu (x) \ne B_{\mu\nu}, \ \ \ U_1,
\ee
%15%%%%12
\be
H'_{\mu\nu}(x) = H_{\mu\nu}(x)+\p_\mu \om_\nu - \p_\nu \om_\mu -if\left[\int^{x}\om_{\ld}(x') dx'^{\ld},  H_{\mu\nu}(x)\right],  \ \ \   SU_3,
\ee
%8%%%%28%%27%%526%%%%%%%%13%%%%9%%%8%%%%16%%%13
\be
\p^\mu B'_{\mu\nu}(x)= \p^\mu B_{\mu\nu}(x) +\p^\mu\p_\mu \Ld_\nu - \p^\mu\p_\nu \Ld_\mu =\p^\mu B_{\mu\nu}(x), \ \ \  U_1,
\ee
%17%%%14
\be
\p^\mu H'_{\mu\nu}(x)= \p^\mu H_{\mu\nu}(x) -if\left[\int^{x}\om_{\ld}(x') dx'^{\ld}, \p^\mu H_{\mu\nu}(x)\right], \ \ \ SU_3,
\ee
%%%%9%%29%%28%%%%27%%%%%%%%%%14%%%%10%%%9%%%18%%%%15
provided the gauge functions $ \Ld_\mu(x)$ and $\om^a_\mu (x)$ satisfy the following constraints
\be
 \p^2 \Ld_\mu(x) - \p_\mu \p^\ld \Ld_\ld(x) = 0,    \ \ \   \p^2=\p_\mu \p_\nu \eta^{\mu\nu},   \ \ \ \ \   \eta^{\mu\nu}=(1,-1,-1,-1),
\ee
%%%%4%%%%%%%%24%%23%%17%21%%%%7%%%4%%%%18%%%%16
\be
\p^\mu \{\p_\mu \om_\nu (x) - \p_\nu \om_\mu (x)\} -if[\om^\mu(x) , H_{\mu\nu}(x)] = 0.
\ee
%%%%%%%%%%%%19%%%%%17
These constraints are necessary for the gauge invariance of the Lagrangians with $U_1$ and $SU_3$ groups.   We stress that the Lagrangian with new $SU_3$ gauge invariance can be constructed only for the general gauge function, $\om_{\mu}(x) \ne \p_{\mu} \om(x)$.   The relation (17) is required to be satisfied for the general case, $\om_{\mu}(x) \ne \p_{\mu} \om(x)$.  The restriction for $\om_\mu(x)$ in (17) is similar to that for gauge functions of Lie groups in the usual non-Abelian gauge theories\cite{9,10}.  We also have 
\be
 \ov{\psi}'(x)\g^\mu \De'_{b\mu} \psi' (x) =  \ov{\psi}(x)\g^\mu \De_{b\mu} \psi(x) , \ \ \ \ \  U_1,
\ee
%%%%%%10%%%30%%29%%%28%%%%%15%%%%11%%%10%%%%%%%19%%%%16%%%18
\be
 \ov{q}'(x)\g^\mu\De'_\mu q'(x) = \ov{q}(x)\g^\mu\De_\mu q(x), \ \ \ \ \ \  SU_3.
\ee
%%20%%%%17%%%%19
Here, we have used the Jacobi identity for the generators $L^a$ and the relations
\be
\p_\mu  \Om(x) =  - ig_b   \Ld_\mu (x) \Om(x),
\ee
%%%%%%%%%%%20
\be
\p_\mu  \Om_{\om}(x) =  - i f  \om^a_\mu (x) L^a \Om_{\om}(x),
\ee
%%%%%%%%%%%%16%%%512%%%11%%%%00000000021
to obtain the results (12)-(17). 

These relations (1), (3), (5), (7) and (9) define generalized $U_1$ gauge symmetry.  There are four components of the gauge vector function $\Ld_\mu$ in (1).  Nevertheless, we impose four constraints (16) for $\Ld_\mu$.  The transformation of $\psi$ in (3) involving a scalar phase factor, and the form of the gauge covariant derivative (10) is the same as that of the $U_1$ gauge group.  Furthermore, in the special case in which the vector function $\Ld_\mu(x)$ can be expressed as the space-time derivative of an arbitrary scalar function $\om(x)$, i.e., $\Ld_{\mu}=\p_{\mu} \om(x)$, the constraint (16) becomes an identity for arbitrary function $\om(x)$.  In this case, the transformations (1), (3) and (5) reduce the usual $U_1$ gauge transformations.  In other words, the non-integrable phase factors become usual phase factors of the $U_1$ group.  Thus, the transformations (1), (3), (5) and  the constraint (16) may be considered as a  generalized $U_1$ gauge transformation.   Similar situation holds in the generalized $SU_3$ gauge transformation.  We shall call them the $U_1$ or $SU_3$ `taiji gauge transformations'\footnote{In ancient Chinese thought, `taiji'  denotes the ultimate principle or the condition that existed before the creation of the world.} to distinguish them from the usual gauge transformations.  Their corresponding gauge fields may be called `taiji gauge fields,' which will satisfy fourth-order partial differential equations, as we shall see below.

\section{A Model of Charmonium and Confining Potential}

Let us apply $SU_3$ taiji gauge symmetry in (11), (15) and (17) to construct a confining model for charmonium.  In this model one has a linear attractive potential between quark and antiquark in a charmonium.   This model suggests a new approach to quark confinement based on the confining potential derived from  the fourth-order gauge field equations dictated by the taiji gauge symmetry.  Thus, the new approach differs   from that in the conventional quantum chromodynamics (QCD) based on the usual gauge symmetry.  This difference will be further discussed in section 4. 

In this model for charmonium, the previous discussions still hold when we replace quark $q(x) $ and anti-quark  $\ov{q}(x)$ by the charmed quarks and antiquarks, $c(x)$ and $\ov{ c}(x)$.   The Lagrangian with $[SU_3]_{color}$ taiji gauge symmetry takes the form
\be
L_{cH}=\frac{1}{2}L_o^2 \p_{\mu} H_{a}^{\mu\ld}\p^{\nu} H_{a\nu\ld} + \ov{c} [i\g^\mu(\p_\mu+ifH^a_{\mu} L^a)- m_c]c,
\ee
%%41%%%%%%%40%%%%18%%%22
where $m_c$ is the mass of the charmed quark\cite{8}.  For the following discussions of classical gauge fields and static potentials, it is not necessary to include a gauge fixing term in the Lagrangian (22), similar to the usual discussion of the Coulomb potential in electrodynamics.   Nevertheless, when one discusses quantum fields and rules for Feynman diagrams, one must include gauge fixing terms such as $L_{gf}= ({L_o^2}/{2\a})(\p_\ld \p_\mu H^{\mu a})(\p^\ld \p^\nu H^a_{\nu})$ in (22), so that the propagator of the massless gauge boson is well defined.

The gauge field equation for $H^a_\mu(x)$ can be derived from the Lagrangian (22),
\be
\p^2 \p^\mu H^a_{\mu\nu} -\frac{f}{L_o^2}\ov{c} \g_\nu L^a c = 0.
\ee
%%12%%34%%%%38%%%19%%%%24%%23
The charmed quark equations are given by
\be
i\g^\mu ( \p_\mu + if H_{a\mu}L_a)c - m_c c= 0, 
\ee
%%%%%%%%13%%%35%%%%39%%%20%%%525%%%%24
\be
i\p_\mu\ov{c} \g^\mu + \ov{c}\g^\mu f H_{\mu a} L_a + m_c \ov{c} = 0.
\ee
%%36%%540%%%%%21%%%%25

Since $H^a_{\mu\nu}$ is anti-symmetric in $\mu$ and $\nu$, the taiji gauge equation (23) implies the continuity equation
\be
\p^\nu(\ov{c} \g_\nu L^a  c )=0,
\ee
%%37%%%5 4 1%%%%22%%%27%%%26
associated with the color  $SU_3$ group.  This continuity equation can also be derived from the charmed quark equations (24) and (25).

The static equation for the time-component  $H^a_0({\bf r})$ in (23) takes the form 
\be
L^2_o \nabla^2 \nabla^2 H^a_0 ({\bf r})= f\ov{c}\g_0 L^a c.
\ee
%%%%%%23%%%%28%%%27
For example, the sources of $H^3_\mu(x)$ and $H^8_\mu(x)$ fields are respectively the color isotopic charge $Q_3$ and the color hypercharge $Q_8$.  All color charges can be expressed in terms of the unit color charge $f$\cite{8}.
  Similar to the static Coulomb potential in electrodynamics, suppose we replace  the time component of the source term $ f( \ov{c}\g_0 L^a c)$ by a point color charge  $Q_3$ at the origin for the component  $a=3$.   We have
\be
L_o^2 \nabla^2 \nabla^2 H^3_0 ({\bf r})=  Q_3 \de^3({\bf r}).
\ee
%%%%%%%%%%%%%45%%%%%35%%34%%%%%%%41%%%%%%24%%%%29%%%%28
It leads to the linear attractive gauge potential fields $H^3_0({\bf r})$ 
\be
H^3_0({\bf r}) = \frac{Q_3}{L^2_o}\frac{ r}{8\pi},  
\ee
%42%%%25%%%%30%%%%29
where we have used the Fourier transform of generalized functions\cite{11,2}.
In addition to this solution, we observe that the fourth-order equation (28) also has a Coulomb-type solution $H'^3_0 ({\bf r}) =b/r$, which satisfies $\nabla^2 H'^3_0 ({\bf r})=0$ for $r \ne 0$.  To determine the unknown constant b, we require that this solution of homogeneous equation satisfies the boundary condition,$ \nabla^2 H'^3_0 (0)=Q_3 \de^3(0)$, at the origin.  This boundary condition implies that the new gauge boson $H^3_\mu$ coupled to other particles with only one single color charge $Q_3$.  Thus we have $b/r = - Q_3/(4\pi r).$
These two solutions, $H^3_0({\bf r})$ and $H'^3_0({\bf r})$, lead to the strong attractive potential energy $Q_3(H^3_0+H'^3_0)=V_{st}$ for the charmonium,
\be
V_{st}=\frac{(Q_3)^2 r}{8 \pi L_o^2} - \frac{(Q_3)^2}{4\pi r}.
\ee
%%%%%%%%%43%%%%26%%%31%30
Besides, there is an attractive electromagnetic potential energy between charmed quark and anti-quark with the electric charges $Q_{e1}=2e/3$ and $Q_{e2}=-2e/3$, we have
\be
V_{em}= \frac{Q_{e1} Q_{e2}}{4\pi r}=  \frac{-4}{9}\frac{e^2}{4\pi r}, \ \ \ \ \ \     \frac{e^2}{4\pi}= \frac{1}{137.035},     
\ee
%%%%%%%%44%%%%27%%%%32%%31
where $c=\hbar=1$.  To estimate the constants $L_o$ and unit color charge $f$ in the Lagrangian (22), we set $Q_3 \approx f$ and use the coefficients of the effective potential energy $V(r)$ obtained by fitting the spectrum of charmonium\cite{1,12},
\be
V(r) = -\frac{\a_c}{r}\left[1-\left(\frac{r}{a}\right)^2 \right],  \ \ \ \  \a_c=0.2, \ \ \ \ \   a=0.2 fm, \ \ \ \   m_c = 1.6  GeV.
\ee
%%%28%%%%33%%%32
  In this approach of Cornell group for charmonium spectrum, heavy charmed quarks are considered as non-relativistic particles.
It is natural to identify the effective potential $V(r)$ in (32) with the sum of the confining potential (30) and the electromagnetic potential (31),
\be
V(r)=V_{st} + V_{em}= \left(\frac{f^2 }{4\pi L_o^2}\right)r  -\left(\frac{f^2}{4\pi} + \frac{e^2}{9\pi}\right)\frac{1}{r}, \ \ \ \  f \approx  Q_3.
\ee
%%%%%%29%%%%%%34%%%%33
Based on the relations (32) and (33), we can estimate the coupling strength $f^2/(4\pi)$ of unit color charge $f$ and the length scale $L_o$:
\be
\frac{f^2}{4\pi} \approx 0.19, \ \ \ \ \ \   L_o \approx 0.93 a = 0.19 fm, \ \ \ \ \    m_c\approx 1.6 GeV. 
\ee
 Since the experimental result for $m_c$ is given by $1.18 GeV \le m_c \le 1.34 GeV$,\cite{12} one may consider  $f^2/(4\pi)$ and $L_o$ in (34) as approximate results with uncertainties of $\approx 25\%$ .

\section{Discussions}

The taiji $U_1$ gauge symmetry was used to discuss the conservation of baryonic charges and baryonic gauge fields, which could provide a field-theoretic understanding of the accelerated cosmic expansion\cite{2,13,14}.  We may remark that the formulation of taiji gauge symmetry for $SU_3$  can be generalized to $SU_N$ without difficulty. 

In the usual formulation of, say, the $U_1$ gauge field $h_\mu$, the  $U_1$ curvature $h_{\mu\nu}=\p_\mu h_\nu -\p_\nu h_\mu$ is gauge invariant.  Thus, $\p^\mu h_{\mu\nu}$,  $\p^\ld h_{\mu\nu}$ etc.  are also $U_1$ gauge invariant, so that one can use them to construct various gauge invariant Lagrangians to obtain higher-order gauge field equations.  But these types of gauge invariant Lagrangians are not unique\cite{2,13}.  Taiji $U_1$ gauge symmetry implies that only the special form $\p^\mu h_{\mu\nu}$ is invariant, as shown in (14).

If one uses the potential (33) to fit the bound states of quark and antiquark systems for both c-quarks and b-quarks, one will get the results $f^2/(4\pi)=0.51, \ L_o=0.34 fm, \ m_c = 1.84 GeV$ and $m_b=5.17 GeV$\cite{15,12}.  Thus, in the low energy regime, the strength of color charge is probably given by $0.19 \le f^2/(4 \pi) \le 0.51$.  Since the value of $m_c=1.84 GeV$ is larger than that associated with (32) or (34), the inclusion of b-quarks seems to give a  less reliable approximate value for $f^2/(4\pi)$, in comparison with that given by (34).

Recently, a full lattice QCD with almost physical quark masses was employed to study charmonium potential.\cite{16,17}  The potentials with spin-independent and spin-dependent parts are obtained by using Bethe-Salpeter wave function in dynamical lattice QCD simulations.  It was found that the spin-independent charmonium potential is quite similar to the non-relativistic potential such as the Cornell potential.\cite{1}  The potential calculated in Bethe-Salpeter amplitude methods exhibits a linear potential at large distances and a Coulomb-type potential at short distances.   But, the Coulomb-type potential may have large uncertainties.  In order to avoid the large discretization error, one uses data which suffer less from the rotational symmetry breaking in the finite cubic box.\cite{16}  The spin-dependent potential obtained in the lattice QCD clearly differs from a repulsive $\delta$-function potential generated by one-gluon exchange, which was widely used in non-relativistic potential model for charmonium.  The r-dependent part in the spin-dependent potential may take the Yukawa form, or Gaussian form, etc.  It is barely consistent with the phenomenological model.\cite{18}

In comparison with the elaborated dynamical lattice QCD simulations, the present confining model suggests a simple understanding and derivation of both the linear and Coulomb-type potential from one single gauge field equation.   This equation is the fourth-order field equation with a generalized gauge symmetry involving non-integrable phase factor in $(SU_3)_{color}$ gauge transformations.  The situation is somewhat similar to that in the electromagnetic field with $U_1$ gauge symmetry.  The spin-dependent potential is not discussed in this confining model and needs further investigation.

It is interesting that the strong coupling strength $0.19 \le f^2/(4 \pi) \le 0.51$ in the gauge covariant derivative (8) and the Lagrangian (22) turns out to be smaller than 1 in the confining model for charmonium.  This result suggests that perturbation calculations could be useful in this confining model.  It also suggests that we could have a satisfactory quantum quark dynamics with strong interaction based on the taiji gauge symmetry, similar to quantum electrodynamics with the usual  $U_1$ gauge symmetry.  Furthermore, the length $L_o= 0.19 fm$ may be used as a reference for large or small quark and antiquark separation in a charmonium.  It appears that the presence of  this length $L_o$  is a general property of the Lagrangian with the taiji gauge symmetry.  It could be considered as the characteristic length in quark dynamics with taiji gauge symmetry.

From the viewpoint of quantum field theory, the degree of divergence in higher order Feynman diagrams associated with the Lagrangian (22) appears to be no worse than that of the corresponding diagram in the usual gauge theory.  Thus, the confining model may be renormalizable.  There is also a linear potential between the c-quark and the new massless gauge boson, which is due to the exchange of a virtual gauge boson between them.  If the massless gauge boson were to escape from a quark-antiquark system, then one will have unphysical massless-particle with negative energies and unphysical runaway solution of gauge fields\cite{5,6,7}.  It is gratifying that these unphysical phenomena cannot be realized because of the confinement property.
 
 In summary, we have demonstrated that by postulating the taiji gauge symmetry we have a new gauge field obeying (23) and (27), we can qualitatively understand the confinement of c- and $\ov{c}$-quarks, and of the massless gauge boson in charmed quark-antiquark system\cite{19}.   The qualitative understanding of confinement can be extended to other quark-antiquark  systems, but its applications to baryonic systems need further study\cite{20,21}.

\bigskip

{\bf ACKNOWLEDGMENTS}

The author would like to thank L. Hsu and D. Fine for useful discussions.  The work was support in part by the Jing Shin Research Fund, UMass Dartmouth Foundation.

\bigskip
Note added.  If we do a  more detailed calculation of the charmonium potential in (33) by considering all source terms, we obtain smaller values,  ${f^2}/{4\pi} \approx 0.04$, and  $L_o \approx  0.013 fm$.

\newpage  
\bibliographystyle{unsrt}

%{99}

\end{document}